\documentclass[aps,pra,reprint,twocolumn,superscriptaddress,showpacs]{revtex4-1}
\usepackage[english]{babel}
\usepackage{float}
\usepackage{amsfonts}
\usepackage{amsmath}
\usepackage{amssymb}
\usepackage{latexsym}
\usepackage{array}
\usepackage{multirow}
\usepackage{ifpdf}
\usepackage{esint}
\usepackage{dsfont}
\usepackage{slashed}
\usepackage{color}
\usepackage{soul}
\usepackage{datetime}

\ifpdf
\usepackage[pdftex]{graphicx}
\else
\usepackage{graphicx}
\fi

\ifpdf \DeclareGraphicsExtensions{.pdf, .jpg, .tif, .png} \else
\DeclareGraphicsExtensions{.eps, .jpg} \fi

\newcommand{\D}{\mathrm{d}}
\newcommand{\WKB}{\mathrm{WKB}}
\newcommand{\dB}{\mathrm{dB}}
\newcommand{\iin}{\mathrm{in}}
\newcommand{\out}{\mathrm{out}}
\newcommand{\lef}{\mathrm{L}}
\newcommand{\rig}{\mathrm{R}}
\newcommand{\npsi}{\varPsi}
\newcommand{\tz}{\tilde{z}}
\newcommand{\tF}{\tilde{F}}
\newcommand{\tk}{\tilde{k}}
\newcommand{\tpsi}{\tilde{\varPsi}}
\newcommand{\tphi}{\tilde{\phi}}
\newcommand{\hz}{\hat{z}}

\newcommand{\bz}{\mathbf{z}}
\newcommand{\bF}{\mathbf{F}}
\newcommand{\bE}{\mathbf{E}}
\newcommand{\bV}{\mathbf{V}}
\newcommand{\bI}{\mathbf{I}}

\newcommand{\bpsi}{\mathbf{\Psi}}
\newcommand{\cS}{\mathcal{S}}
\newcommand{\cT}{\mathcal{T}}

\newcommand{\cW}{\mathcal{W}}
\newcommand{\cI}{\mathcal{I}}
\newcommand{\Eunit}{\mathrm{E}_1}
\newcommand{\hunit}{\mathrm{h}_1}

\begin{document}

\title{Liouville transformations and quantum reflection}

\author{G. Dufour} \email[]{gabriel.dufour@upmc.fr}
\affiliation{Laboratoire Kastler Brossel, UPMC-Sorbonne
Universit\'es, CNRS, ENS-PSL Research University, Coll\`ege de
France, Campus Jussieu, F-75252 Paris, France.}
\author{R. Gu\'erout}
\affiliation{Laboratoire Kastler Brossel, UPMC-Sorbonne
Universit\'es, CNRS, ENS-PSL Research University, Coll\`ege de
France, Campus Jussieu, F-75252 Paris, France.}
\author{A. Lambrecht}
\affiliation{Laboratoire Kastler Brossel, UPMC-Sorbonne
Universit\'es, CNRS, ENS-PSL Research University, Coll\`ege de
France, Campus Jussieu, F-75252 Paris, France.}
\author{S. Reynaud}
\affiliation{Laboratoire Kastler Brossel, UPMC-Sorbonne
Universit\'es, CNRS, ENS-PSL Research University, Coll\`ege de
France, Campus Jussieu, F-75252 Paris, France.}
\date{liouvilletrans.tex, \today, \currenttime}

\begin{abstract}
Liouville transformations of Schr\"odinger equations preserve the
scattering amplitudes while changing the effective potential. We
discuss the properties of these gauge transformations and introduce
a special Liouville gauge which allows one to map the problem of
quantum reflection of an atom on an attractive Casimir-Polder
\emph{well} into that of reflection on a repulsive \emph{wall}. We
deduce a quantitative evaluation of quantum reflection probabilities
in terms of the universal probability which corresponds to the
solution of the $V_4=-C_4/z^4$ far-end Casimir-Polder potential.
\end{abstract}

\maketitle

\section{Introduction}

Quantum reflection of atoms from the van der Waals attraction to a
surface has been studied theoretically since the early days of
quantum mechanics \cite{Lennard-Jones1936,Lennard-Jones1936a}.
Though the classical motion would be increasingly accelerated
towards the surface, the quantum matter waves are reflected back
with a probability that approaches unity at low energies, due to
the rapid variation of the potential close to the surface.

Quantum reflection was first studied experimentally for He and H
atoms on liquid helium films \cite{Nayak1983,Berkhout1989,Yu1993}
and more recently for ultracold atoms or molecules on solid surfaces
\cite{Shimizu2001,Druzhinina2003,Pasquini2004,Oberst2005,Pasquini2006,%
Zhao2010,Zhao2011}. Meanwhile, many theoretical papers have studied
various fundamental aspects and applications of quantum reflection
\cite{Berry1972,Boheim1982,Clougherty1992,Carraro1992,Henkel1996,%
Friedrich2002,Friedrich2004,Friedrich2004a,Judd2011}. More recently,
it has been noticed that quantum reflection could be useful for
storing and guiding cold antihydrogen atoms
\cite{Voronin2005pra,Voronin2005,Voronin2011,Voronin2012} and that
it should play a role in any experiment where antihydrogen atoms
interact with a matter plate (see
\cite{Dufour2013qrefl,Dufour2013porous,Dufour2014} and references
therein).

Paradoxical results appear in the study of quantum reflection (QR)
from the Casimir-Polder (CP) interaction. The probability of quantum
reflection not only increases when the velocity of the incident atom
is decreased, but also when the magnitude of the interaction is
decreased. For example, the probability of quantum reflection is
larger for atoms falling onto silica bulk than onto metallic mirrors
\cite{Dufour2013qrefl} and is even larger for nanoporous silica
\cite{Dufour2013porous}. This paradox is qualitatively explained by
the fact that atoms get closer to the surface for a weaker
potential, so that the CP potential becomes steeper and quantum
reflection probability larger.

In the present paper, we propose a quantitative treatment of these
paradoxical behaviors based on Liouville transformations. Such
transformations are gauge transformations of the Schr\"odinger
equation, which map problems corresponding to different potentials
into one another, while leaving scattering amplitudes invariant. In
the case of QR on a CP potential studied in this paper, a special
Liouville gauge can be introduced to transform the potential from an
attractive CP \emph{well} into a repulsive \emph{wall}. The
paradoxical features of the initial QR problem become intuitive
predictions of the transformed problem. Furthermore, QR
probabilities can then be described in terms of the universal
solution associated with the $V_4=-C_4/z^4$ far-end Casimir-Polder
potential.

In \S~\ref{sec:quantum}, we recall the usual treatment of the QR
problem, based on deviations from the semiclassical WKB
approximation \cite{Berry1972}. We present in \S~\ref{sec:liouville}
the Liouville transformations which transform Schr\"odinger
equations into equivalent ones corresponding to different
potentials. We then introduce in \S~\ref{sec:special} a special
choice which maps the original problem of QR on a CP well into a
more intuitively understood problem of reflection on a repulsive
wall. In \S~\ref{sec:C4model} we study the $V_4=-C_4/z^4$ potential
which shows non trivial symmetry properties while being
representative of the CP interaction in the far-end. We finally use
these results (\S~\ref{sec:discussion}) to give a simple evaluation
of QR probabilities, in terms of the universal function associated
with this problem and of one scattering length parameter depending
on the full CP potential.

\section{Quantum reflection}
\label{sec:quantum}

We consider a cold atom of mass $m$ incident with a velocity $v<0$
parallel to the $z-$axis upon the CP potential $V(z)$ in the
half-space $z>0$ above the material surface located at $z=0$. We
consider a plane material surface, so that the vertical motion is
decoupled from the horizontal one.

The vertical motion is then described by a 1D Schr\"odinger
equation:
\begin{eqnarray}
\label{schrod} &&\npsi^{\prime\prime}(z) + F(z) \, \npsi(z) = 0~,\\
&& F(z)\equiv\frac{2m\left(E-V(z)\right)}{\hbar^2}~.
\end{eqnarray}
The primes represent the derivative of a function with respect to
its argument. The CP potential $V(z)$ is attractive, with
characteristic inverse power laws at both ends of the $z-$domain,
the \emph{cliff-side} close to the surface $z\to0$ and the
\emph{far-end} $z\to\infty$:
\begin{eqnarray}
\label{powerlaws}
&&V(z) \underset{z\to0}{\simeq} V_3(z) ~,\quad V_3(z)\equiv -C_3/z^3~,\\
&& V(z) \underset{z\to\infty}{\simeq} V_4(z) ~,\quad V_4(z)\equiv
-C_4/z^4 ~.
\end{eqnarray}

In \eqref{schrod}, $E=\tfrac12 mv^2$ is the energy associated with
the motion orthogonal to the plane. It may be the result of a free
fall from a height $h$ above the surface $E=mgh$, where $g$ is the
acceleration of gravity. This semi-classical treatment of the free
fall is possible when $E$ is much larger than the energy $\Eunit$ of
the first gravitational quantum state above the surface
\cite{Voronin2006,Dufour2014}:
\begin{equation}
\label{Eunit} \Eunit \equiv \left(\frac{\hbar^2 m g^2}2
\right)^{\frac13} \lambda_1 \simeq 1.407 \ \mathrm{peV} ~,
\end{equation}
where $\lambda_1\simeq2.338$ is the absolute value of the first zero
of the Airy function Ai. In the following, we use $\Eunit$ as the
unit for $E$ so that the validity of the semi-classical treatment of
the free fall above the surface is simply $E/\Eunit\gg1$. This
condition also corresponds to $h/\hunit\gg1$ where
$\hunit\equiv\Eunit/mg$ is the associated unit for the free fall
height $h$, that is the height $\simeq13.7\mu$m of the classical
turning point for the first quantum state \cite{Voronin2006}.

If we were treating also the interaction with the CP potential in a
semiclassical approach, the function $F(z)$ would be seen as the
square of the de Broglie wave-vector $k_\dB$ associated with the
classical momentum $\hbar k_\dB$:
\begin{equation}
\label{wkbk} k_\dB(z) \equiv \sqrt{F(z)}~,
\end{equation}
As the CP potential is attractive and the incident energy positive,
$F$ is everywhere positive, so that a classical particle undergoes
an increasing acceleration towards the surface. This behavior is
mimicked by the semiclassical WKB wave-functions which propagate in
the rightward and leftward directions ($\eta=+1$ and $-1$
respectively):
\begin{eqnarray}
\label{wkb} &&\npsi_\WKB ^\eta(z) = \alpha_\dB(z) \, e^{i\eta
\phi_\dB(z)} ~,\\ \label{wkbphi} &&\alpha_\dB(z) =
\frac1{\sqrt{k_\dB(z)}}~, \quad \phi_\dB(z) = \int_{z_0}^z k_\dB(z')
\D z'~.
\end{eqnarray}
Here, $\alpha_\dB$ is the WKB amplitude and $\phi_\dB$ the WKB phase
associated with the classical action integral $\hbar\phi_\dB$. The
arbitrariness of the choice of the reference point $z_0$ is fixed in
the following by the far-end convention:
\begin{eqnarray}
\label{convz0} \underset{z\to\infty}{\lim} \left(\phi_\dB(z)-\kappa
z\right) = 0~,\\
\label{defkappa} \kappa = \underset{z\to\infty}{\lim} k_\dB(z) =
\frac{\sqrt{2mE}}\hbar~.
\end{eqnarray}

In a quantum treatment of the interaction with the surface
\cite{Dufour2013qrefl}, QR appears as a consequence of non-adiabatic
transitions between the counter-propagating WKB waves \eqref{wkb}.
It can be obtained by solving the exact Schr\"odinger equation
\eqref{schrod} for a wave-function written, in full generality, as a
linear combination of $\npsi_\WKB ^\eta(z)$ with $z-$dependent
coefficients $\beta_\eta$:
\begin{equation}
\label{bpm} \npsi(z) = \beta_\eta(z) \npsi_\WKB^\eta(z) ~,
\end{equation}
where we use an implicit sum rule for repeated indices. These
coefficients obey coupled first-order differential equations
\cite{Berry1972}:
\begin{equation}
\label{coupled} \beta_\eta^\prime(z) = \beta_{-\eta}(z) \,
\frac{k_\dB^\prime(z)}{2 k_\dB(z)} e^{- 2i \eta \phi_\dB(z)} ~.
\end{equation}
This system of equations can be solved numerically and matched to
the WKB solutions at both ends of the domain
\cite{Cote1997,Mody2001,Friedrich2002}. Special care has to be taken
on the cliff-side where the potential diverges. The matching has to
use the mathematical solutions of \eqref{schrod} known for the $V_3$
potential \cite{Voronin2005pra,Dufour2013qrefl}, at the price of
losing physical understanding of the problem.

Matter-waves can be reflected back from the cliff-side so that the
complete problem depends on the details of the physics of the
surface, including possible sticking, non specular reflection or
annihilation for antimatter. In this letter, we focus our attention
on the one-way problem where the CP potential is crossed only once
and, therefore, do not discuss this surface physics problem any
longer. The numerical solution of \eqref{schrod} leads to reflection
and transmission amplitudes depending on the incident energy $E$ or,
equivalently, on the parameter $\kappa$ defined in \eqref{defkappa}.
A qualitative criterion for occurrence of QR is that the coupling
term in \eqref{coupled} takes significant values, which may be
stated as a large enough variation of the de Broglie wavelength on
the length scale fixed by the de Broglie wavelength.

A less approximate discussion of this point can be based on the
remark that WKB wave-functions obey an equation differing from the
original one \eqref{schrod}:
\begin{eqnarray}
\label{schrodwkb} \npsi_\WKB^{\eta\;\;\prime\prime}(z) + F(z) \,
\npsi_\WKB^\eta(z) = \frac{\alpha_\dB''(z)}{\alpha_\dB(z)}
\npsi_\WKB^\eta(z) ~.&&
\end{eqnarray}
The difference between \eqref{schrod} and \eqref{schrodwkb} is often
described in terms of the so-called \emph{badlands function}
\cite{Dufour2013qrefl}:
\begin{eqnarray}
\label{badlandsalpha} Q(z) = -
\frac{\alpha_\dB''(z)}{F(z)\alpha_\dB(z)} = - \alpha_\dB^3 (z)
\alpha_\dB''(z) ~, &&
\end{eqnarray}
which can also be expressed in terms of the Schwarzian derivative
$\{\phi_\dB,z\}$ of the WKB phase $\phi_\dB$:
\begin{eqnarray}
\label{badlands} Q(z) = \frac{\{\phi_\dB,z\}}{2 F(z)} =
\frac{\{\phi_\dB,z\}} {2 k_\dB^2(z)}  ~.&&
\end{eqnarray}
The Schwarzian derivative is defined for $f(z)$ as:
\begin{eqnarray}
\label{schwarz} &&\{f,z\} = \frac{f'''(z)}{f'(z)} - \frac{3}{2}
\frac{f''(z)^2}{f'(z)^2} ~.
\end{eqnarray}

\begin{figure}[ht]
  \begin{center}
    \includegraphics[width=0.5\textwidth]{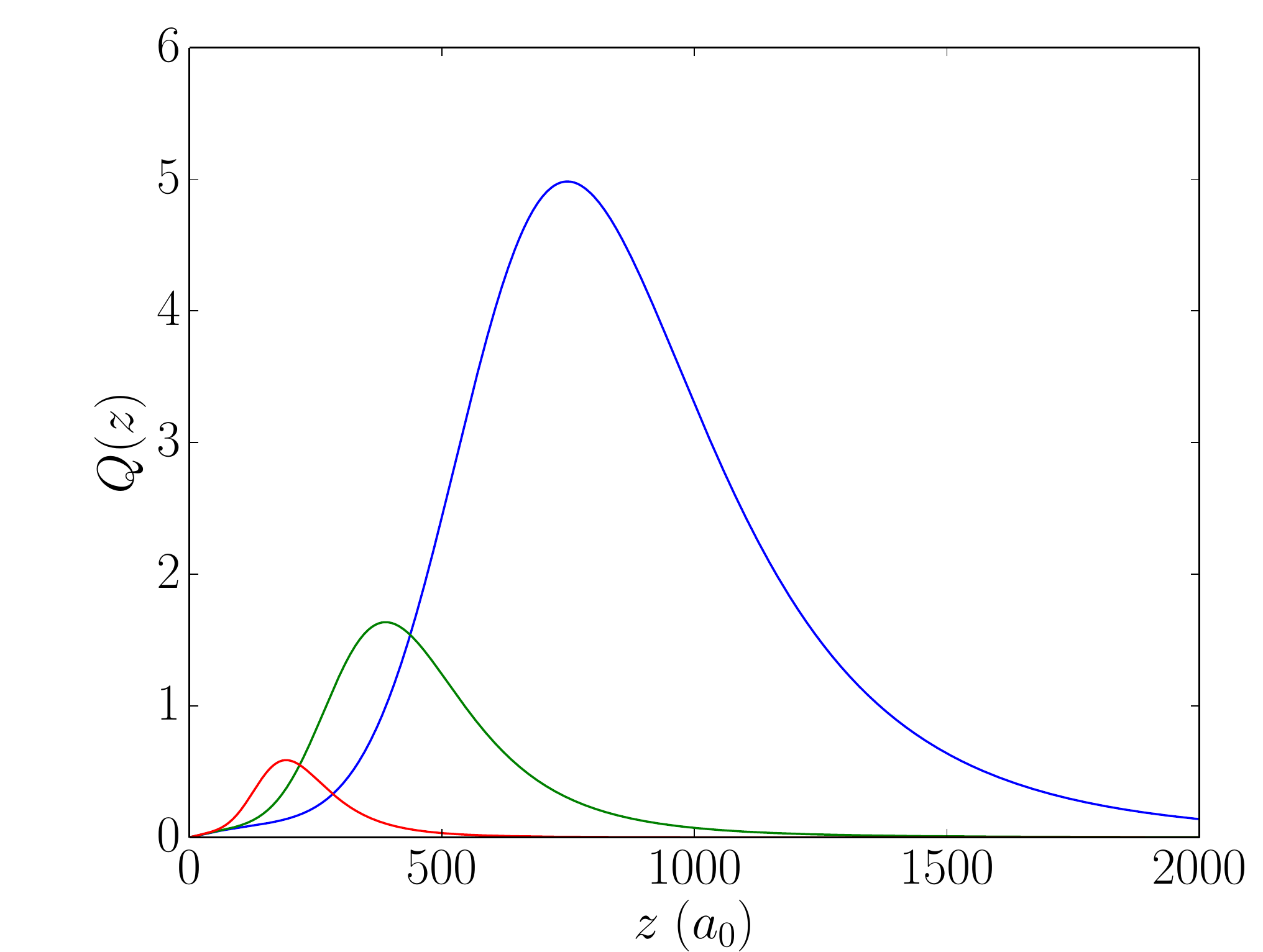}
  \end{center}
\caption{[Colors online] The blue, green and red curves (from the
highest to the lowest peak) show the functions $Q(z)$ calculated for
energies $E$ respectively equal to $10^3$, $10^4$ and $10^5 \Eunit$
and $V(z)$ calculated for an hydrogen atom and a silica bulk.
\label{fig:badlands} }
\end{figure}

An important property of the function $Q(z)$ is that it vanishes not
only in the far-end where $k_\dB$ goes to a constant, but also at
the cliff-side as a consequence of the power-law variation
\eqref{powerlaws} of the Casimir-Polder potential. It follows that
the regions where QR takes place are indicated by significant values
of the peaked function $Q(z)$. This property has been proven by
several examples in \cite{Dufour2013qrefl,Dufour2013porous} and it
is also illustrated on Fig.\ref{fig:badlands}.

The three curves show the functions $Q(z)$ for an hydrogen atom
above a silica bulk with energies $E$ respectively equal to $10^3$,
$10^4$ and $10^5 \Eunit$. The curves on Fig.\ref{fig:badlands} have
a higher and higher peak for lower and lower energies, with a peak
farther and farther from the cliff. This variation of the peak value
is well correlated with the QR probability calculated by solving the
Schr\"odinger equation as indicated previously (details in
\cite{Dufour2013qrefl}). The QR probabilities $R$, given in Table
\ref{QRproba}, indeed increase when the peak values increases, that
is also when the energy decreases.

\begin{table}[H]
\begin{center}
\begin{tabular}{|c|c|c|c|c|c| c|c|}
\hline $E$ [$\Eunit  $] & 1 &10 & $10^2$ & $10^3$& $10^4$ & $10^5$ \\
\hline $R$ [\%] & 98.5 & 95.4 &86.1 & 63.2 &28.0 & 5.6 \\
\hline
\end{tabular}
\caption{Quantum reflection probabilities $R$ for hydrogen atoms
falling on a silica bulk plate, with different incident energies
$E$, given in units of $\Eunit$ (see \eqref{Eunit}). The three last
columns correspond to the plots on Fig.\ref{fig:badlands}, while the
first one is given only as an indication of the trend of QR
probability for the lowest gravitational quantum state.
\label{QRproba} }
\end{center}
\end{table}

In spite of its effectiveness, the method reminded in this section
suffers several drawbacks. First QR is a \emph{scattering process}
with incident matter waves reflected or transmitted when crossing
the badlands, but this scattering problem is poorly defined as the
potential diverges at the cliff. Second the correlation of the peak
value of $Q$ with the QR probability is observed, but a quantitative
interpretation of this correlation is missing. In fact, the role of
the badlands function is discussed by comparing the two different
problems associated with \eqref{schrod} and \eqref{schrodwkb}, and
not by studying the equation of physical interest \eqref{schrod}.
All these drawbacks are cured in the sequel of this paper, thanks to
the introduction of Liouville transformations of the Schr\"odinger
equation \cite{Dufour2014liou}.

\section{Liouville transformations}
\label{sec:liouville}

The Schr\"odinger equation \eqref{schrod} is an example of a
Sturm-Liouville equation under Liouville normal form
\cite{Liouville1836}. What is called the WKB approximation by
physicists was in fact introduced by Liouville \cite{Liouville1837}
and Green \cite{Green1838} for studying properties of such
equations, long before Wentzel \cite{Wentzel1926}, Kramers
\cite{Kramers1926} and Brillouin \cite{Brillouin1926} used it in the
context of quantum mechanics. The transformations used in the
present paper were introduced by Liouville in 1837
\cite{Liouville1837} and we follow here the convention of Olver
\cite{Olver1997,Olver2010,dlmf} for naming them after Liouville (see
the notes at the end of ch.~6 in \cite{Olver1997}).

Liouville transformations have been used to obtain approximate
solutions \cite{Langer1931,Miller1953,Dingle1956}. They have been
also used to study second-order differential equations applied to
solvable Schr\"odinger equations
\cite{Bose1964,Natanzon1971,Everitt1982,Milson1998,Derezinski2011}.
We want to emphasize at this point that we use here the Liouville
transformations to describe equivalent scattering problems
corresponding to different potentials, with no approximation, for
potentials not belonging to a class of solvable problems.

After these historical remarks, we define the Liouville
transformations \cite{Olver2010} which correspond to coordinate
changes with correlated rescalings of the wave-function. The
coordinate change maps the physical $z-$domain into a $\tz-$domain
with $\tz(z)$ a smooth monotonous function ($\tz'(z)>0$). Equation
\eqref{schrod} for $\npsi(z)$ keeps the same form for the rescaled
wave-function $\tpsi(\tz)$:
\begin{eqnarray}\label{liouville}
&&\tpsi(\tz)=\sqrt{\tz'(z)}\,\npsi(z) ~, \\
\label{transformedPsi} &&\tpsi''(\tz) + \tF(\tz) \, \tpsi(\tz)=0~,
\end{eqnarray}
with a transformed function $\tF(\tz)$:
\begin{eqnarray}
\label{transformedF}
&&\tF(\tz)=\frac{F(z)-\tfrac{1}{2}\{\tz,z\}}{\tz'(z)^2}~.
\end{eqnarray}
The curly braces denote the Schwarzian derivative \eqref{schwarz} of
the coordinate transformation $\tz(z)$.

The composition of two Liouville transformations  $z\to \tz$ and
$\tz\to\hz$ is a Liouville transformation  $z\to \hz$, and the group
properties of this composition law is ensured by Cayley's identity
for Schwarzian derivatives \cite{Olver2010}:
\begin{equation} \label{cayley}
\left\{\hz,z\right\} =
\left(\tz'(z)\right)^{2}\,\left\{\hz,\tz\right\} +
\left\{\tz,z\right\}.
\end{equation}
When this identity is applied to inverse transformations ($\hz=z$),
the following relation is obtained:
\begin{equation}\label{inverse}
0 = \left(\tz'(z)\right)^{2}\,\left\{z,\tz\right\} +
\left\{\tz,z\right\},
\end{equation}
so that the transformation \eqref{transformedF} can also be written:
\begin{equation}\label{Fexpl}
\tF(\tz)=z'(\tz)^2 F(z) + \tfrac{1}{2} \{z,\tz\}~.
\end{equation}

The Wronskian of two solutions $\npsi_1,\npsi_2$ of the
Schr\"odinger equation is a constant, independent of $z$ and
antisymmetric in the exchange of the two solutions:
\begin{equation}\label{wronskian}
\cW\left(\npsi_1,\npsi_2\right) = \npsi_1(z)\npsi_2'(z) -
\npsi_1'(z)\npsi_2(z) ~.
\end{equation}
The Liouville transformations preserve this Wronskian:
\begin{equation}\label{wronskianinv}
\cW(\npsi_1,\npsi_2) = \tilde\cW(\tpsi_1,\tpsi_2)~,
\end{equation}
and this property has important physical consequences, as shown in
the sequel of this section.

When $\npsi$ is a solution of \eqref{schrod}, its complex conjugate
$\npsi^*$ is also a solution, and the probability density current
$j(z)$ is proportional to the Wronskian of $\npsi^*$ and $\npsi$:
\begin{equation}
\label{defj} j(z) \equiv \frac\hbar{2 i m} \cW(\npsi^*,\npsi)~.
\end{equation}
As the probability density $\rho=\npsi^*\npsi$ is time-independent
in the problem studied in this paper, $j(z)$ is a constant $j$
independent of $z$. It follows from \eqref{wronskianinv} that this
constant is invariant under Liouville transformations $j = \tilde
j$. We emphasize that the probability density $\rho$ is neither
constant nor preserved by the transformation, as one deduces from
\eqref{liouville} that $\rho(z)=z'(\tz)\,\tilde \rho(\tz)$. We also
note that the WKB functions, which are exact solutions of
\eqref{schrodwkb} and approximate solutions of \eqref{schrod} at
both ends of the $z-$domain, obey the following relations (which
will be used below):
\begin{eqnarray}
\label{WKBrel} \cW\left(\left(\npsi^\eta_\WKB\right)^*,
\npsi^{\eta'}_\WKB\right) = {2} \, \cI_\eta^{\eta'} ~,~ \cI =
\begin{pmatrix} i & 0 \\ 0 & -i \end{pmatrix} ~. &&
\end{eqnarray}

\begin{figure}[ht]
  \begin{center}
    \includegraphics[width=0.5\textwidth]{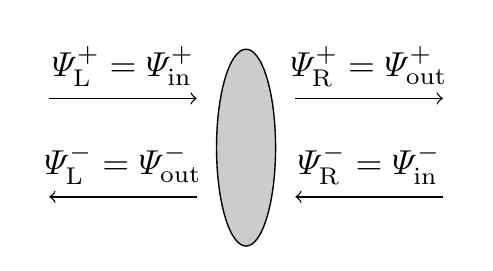}
  \end{center}
\caption{Schematic representation of the four
solutions which match WKB waves on the left and right
sides of the interaction region
symbolized by the grey ellipse.
\label{fig:scattering} }
\end{figure}

We recall now that the scattering amplitudes can be written in terms
of Wronskians of solutions of the Schr\"odinger equation
\cite{Whitton1973}. To this aim, we note that $Q(z)$ goes to 0 at
the left and right ends of the $z-$domain, so that we can define
exact solutions $\npsi^\eta_\lef$ and $\npsi^\eta_\rig$ of
\eqref{schrod} which match the asymptotic WKB waves there:
\begin{eqnarray}
\npsi_\lef^\eta(z) \underset{z\to0}{\to} \npsi_\WKB^\eta(z) ~,\quad
\npsi_\rig^\eta(z) \underset{z\to\infty}{\to} \npsi_\WKB^\eta(z)
 ~. &&
\end{eqnarray}
These four solutions are schematized in Fig.\ref{fig:scattering}.

As the WKB waves $\npsi_\WKB^\eta$ and $\npsi_\WKB^{-\eta}$ are
complex conjugates of each other, this is also the case for the
exact solutions matching them at left or right ends:
\begin{eqnarray}
\left(\psi_\lef^{\eta}\right)^* = \psi_\lef^{-\eta} ~, \quad
\left(\psi_\rig^{\eta}\right)^* = \psi_\rig^{-\eta} ~.
\end{eqnarray}

A generic solution $\npsi(z)$ of \eqref{schrod} can then be
decomposed over the left or right basis:
\begin{eqnarray}
\label{generic} \npsi(z) = a ^\lef_\eta \, \npsi_\lef^\eta(z) = a
^\rig_\eta \, \npsi_\rig^\eta(z) ~,
\end{eqnarray}
For each decomposition, the amplitudes can be collected in column
matrices related by a transfer matrix:
\begin{eqnarray}
\label{Tmatrix} a^\lef_\eta = \cT_\eta^{\eta'} \, a^\rig_{\eta'}
\quad \leftrightarrow \quad \npsi_\rig^\eta = \cT_{\eta'}^\eta \,
\npsi_\lef^{\eta'} ~.
\end{eqnarray}

These solutions and associated amplitudes can alternatively be
defined in terms of \emph{outgoing} and \emph{incoming} waves with
the usual identification:
\begin{eqnarray}
\begin{pmatrix} \npsi_\out^+ \\
\npsi_\out^- \end{pmatrix} = \begin{pmatrix} \npsi_\rig^+ \\
\npsi_\lef^- \end{pmatrix} ~,\quad  \begin{pmatrix} \npsi_\iin^+ \\
\npsi_\iin^- \end{pmatrix} = \begin{pmatrix} \npsi_\lef^+ \\
\npsi_\rig^- \end{pmatrix}  ~. &&
\end{eqnarray}
The \emph{out} and \emph{in} amplitudes are related by the unitary
scattering matrix ($\cS\cS^\dagger=\cI$):
\begin{eqnarray}
\label{Smatrix} a^\out_\eta = \cS_\eta^{\eta'} \, a^\iin_{\eta'} ~,
\end{eqnarray}
which can be obtained from the transfer matrix (see
\cite{Genet2003}):
\begin{eqnarray}
\label{SandT} &&\cS = \begin{pmatrix} \cS_+^+ & \cS_+^- \\
\cS_-^+ & \cS_-^- \end{pmatrix}
=  \frac1{\cT_+^+}
\begin{pmatrix}
 1 & -\cT_+^- \\
 \cT_-^+  & 1
\end{pmatrix}  ~, \\
&&\cT = \begin{pmatrix} \cT_+^+ & \cT_+^- \\ \cT_-^+ & \cT_-^-
\end{pmatrix} ~,\quad \det\cT=1 ~.
\end{eqnarray}

As the Wronskian of solutions $\npsi^\eta_\lef$ or $\npsi^\eta_\rig$
is a constant, it can be evaluated in particular in the asymptotic
regions where these exact solutions reduce to WKB waves. They
therefore obey the same relations as in \eqref{WKBrel}:
\begin{eqnarray}\label{Wrel}
\cW\left(\left(\npsi^\eta_\lef\right)^*,\npsi^{\eta'}_\lef\right) =
\cW\left(\left(\npsi^\eta_\rig\right)^*,\npsi^{\eta'}_\rig\right) =
{2} \, \cI_\eta^{\eta'} .&&
\end{eqnarray}
The information on the scattering is then contained in the
Wronskians involving solutions at the left and right ends:
\begin{eqnarray}
\label{WandT}
\cW\left(\left(\npsi^\eta_\lef\right)^*,\npsi^{\eta'}_\rig\right) =
2 \, \left(\cI\cT\right)_\eta^{\eta'}.&&
\end{eqnarray}

Since the matrices $\cT$ and $\cS$ are expressed in terms of
Wronskians by using \eqref{WandT} and \eqref{SandT}, it follows from
\eqref{wronskianinv} that they are invariant under Liouville
transformations ($\tilde\cT=\cT$ and $\tilde\cS=\cS$). In
particular, the reflection and transmission amplitudes $r=\cS_+^-$
and $t=\cS_-^-$ defined for waves incoming from the far-end
amplitudes are preserved ($r = \tilde r$ and $t = \tilde t$) and can
be calculated equivalently after any Liouville transformations.

It is worth stressing again that these gauge transformations relate
equivalent scattering problems to one another, while not necessarily
making the resolution simpler. In specific cases, for example the
model studied in \S~\ref{sec:C4model}, they may lead to non trivial
\emph{symmetry properties}. In the general case, we show in the next
section that a special gauge choice brings satisfactory answers to
all points raised at the end of \S~\ref{sec:quantum}.

\section{Special gauge choice}
\label{sec:special}

In this section, we choose a special Liouville gauge which shows
interesting properties. Precisely, we choose a coordinate $\bz$
proportional to the WKB phase $\phi_\dB$ which maps the initial
problem of QR on an attractive well into a different problem of
reflection on a repulsive wall. This special gauge choice brings
satisfactory answers to all questions raised above. In particular,
it leads to a perfectly well-defined scattering problem with no
interaction in the asymptotic states, and it also allows to
understand the variation of the QR probability in a rigorous as well
as intuitive manner.

The special gauge choice is fixed by the following definition of the
coordinate $\bz$ and associated quantities identified by boldfacing:
\begin{eqnarray}
\label{specific} && \bz\equiv \frac{\phi_\dB(z)}{\varkappa} \;,
\quad
\bpsi(\bz)=\sqrt{\bz'(z)}\,\npsi(z) ~, \\
&& \bF(\bz)=\frac{F(z)-\tfrac{1}{2}\{\bz,z\}}{\bz'(z)^2} =
\varkappa^2 \left(1- Q(z)\right) ~, \label{specificF}
\end{eqnarray}
where we have noticed that $\{\bz,z\}=\{\phi_\dB,z\}$ and then used
the definition \eqref{badlands} of $Q(z)$. The scale constant
$\varkappa$ is arbitrary at this point, but will be fixed soon.
Equation \eqref{specificF} can be rewritten in terms of energy and
potential:
\begin{eqnarray} \label{specificEV}
\bF(\bz) \equiv \bE - \bV(\bz) \;, \; \bE = \varkappa^2 \;, \;
\bV(\bz) = \varkappa^2 Q(z)  && ~.
\end{eqnarray}
As $Q(z)$ goes to zero at both ends of the physical domain
$z\in\,]0,\infty[$, the interaction potential $\bV$ tends to 0 at
both ends of the transformed domain $\bz\in\,]-\infty,\infty[$. It
thus corresponds to a well-defined scattering problem with no
interaction in the asymptotic input and output states.

Using the expression \eqref{badlandsalpha} of $Q(z)$, a positivity
property can be demonstrated for the integral of $\bV$:
\begin{eqnarray}
\label{integral} \bI &\equiv& \int_{-\infty}^\infty \bV(\bz) \D \bz
= -\varkappa \int_0^\infty \alpha_\dB(z) \alpha_\dB''(z)
\D z \nonumber \\
&=& \varkappa \int_0^\infty \left(\alpha_\dB'(z) \right)^2 \D z > 0
~,
\end{eqnarray}
where the integrated term, which should appear in the integration by
parts between the first and second lines, vanishes at left and right
ends ($\alpha_\dB \alpha_\dB'\to0$ for $z\to0$ or $z\to\infty$).
Whereas the initial potential $V$ was everywhere negative, the shape
of the transformed potential $\bV$ is mostly a repulsive wall, even
though $\bV$ may be negative in some parts of the $\bz-$domain. We
will see now that the transformed equation has classical turning
points where $\bF=0$ or $\bE=\bV$, for not too large values of the
original energy $E$.

Before going further in the discussion of this point, we fix the
scale constant $\varkappa$ to be determined by the wavevector
$\kappa$ and the length scale $\ell$ associated with the far-end
tail of the CP potential:
\begin{eqnarray}
\label{fixscale} \varkappa = \sqrt{\kappa\ell} ~,\quad \kappa =
\frac{\sqrt{2mE}}\hbar ~,\quad \ell = \frac{\sqrt{2mC_4}}\hbar ~.
\end{eqnarray}
This choice will lead to functions $\bV(\bz)$ having nearly
identical peak shapes for different energies $E$, at least for not
too large values of $E$. In fact, these functions reproduce a
universal function $\bV_4(\bz)$ when the initial potential $V(z)$
matches the form of the far-end tail $V_4(z)$  of the CP potential.
This model is studied in \S~\ref{sec:C4model} and it is shown there
that the universal function $\bV_4(\bz)$ has a peak value at
$z=\zeta$ where:
\begin{equation}
\label{invscale} \zeta = \sqrt{\frac\ell\kappa} =
\sqrt[4]{\frac{C_4}E} ~.
\end{equation}

\begin{figure}[ht]
  \begin{center}
    \includegraphics[width=0.5\textwidth]{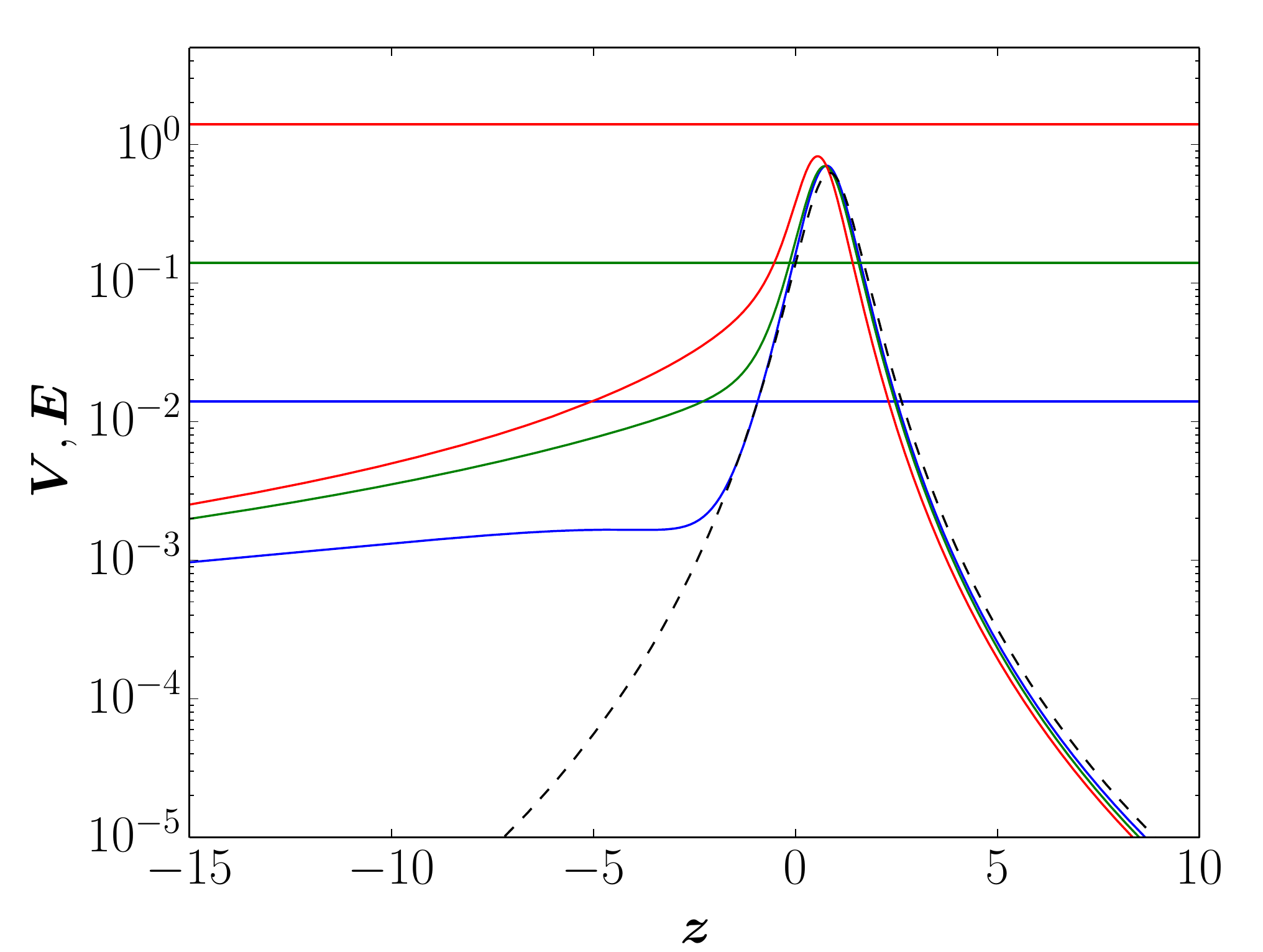}
  \end{center}
\caption{[Colors online] The plots represent the constants $\bE$
(horizontal lines) and the functions $\bV(\bz)$ (curves) calculated
for different scattering problems, corresponding to the same CP
potential $V(z)$ between an hydrogen atom and a silica bulk and
energies $E=10,~ 10^3,~ 10^5 \Eunit$ respectively for the blue,
green and red curves (from the lowest to the highest value of $\bE$,
or from the lowest to the highest value of $\bV$ in the left-hand
part of the plot). The black (dashed) curve is the universal
function $\bV_4(\bz)$ calculated for a $V_4$ model.
\label{fig:silica-energy} }
\end{figure}

The plots on Fig.\ref{fig:silica-energy} show $\bE$ and $\bV$ for CP
potentials $V$ calculated between a hydrogen atom and a silica bulk
\cite{Dufour2013qrefl} and 3 incident energies $E$ respectively
equal to 10$\Eunit$, 10$^3\Eunit$ and 10$^5\Eunit$. The transformed
energies $\bE$ are represented as the 3 horizontal lines and the
transformed potentials $\bV(\bz)$ as the 3 curves. With $\bE$ always
positive and $\bV(\bz)$ mostly positive, a logarithmic scale is used
along the vertical axis, in order to make some details more
apparent.

We see on Fig.\ref{fig:silica-energy} that the peaks for the
functions $\bV$ are nearly the same for the different initial
energies. This is due to the fact that, for the parameters chosen
here, these peaks correspond to distances $z\sim\zeta$ such that the
exact CP potential $V(z)$ is close to its far-end tail $V_4(z)$. The
deviations appearing on the plots correspond to values of $z$ closer
to the cliff-side, where $V_4(z)$ is indeed a poor representation of
$V(z)$. As $\bV$ is nearly the same for the different problems
whereas $\bE=\varkappa^2=\kappa\ell$, it follows that classical
turning points appear in the transformed problem for not too large
energies $E\propto\kappa^2$. For the plots drawn on
Fig.\ref{fig:silica-energy} turning points appear for $E=10,~ 10^3
\Eunit$, but not for $E=10^5 \Eunit$.

The existence of classical turning points in the transformed problem
is in striking contrast with the initial problem of QR on an
attractive well, which did not show turning points. This initial QR
problem has been transformed into the more intuitive problem of
ordinary reflection on a repulsive wall, with exactly identical
scattering amplitudes. The fact that the QR probability goes to
unity when $\kappa\to0$ is now understood as an immediate
consequence of the increasing reflection expected for a particle
with a lower and lower energy $\bE$ coming onto a repulsive wall
with a more or less fixed peak value.

\begin{figure}[H]
  \begin{center}
    \includegraphics[width=0.5\textwidth]{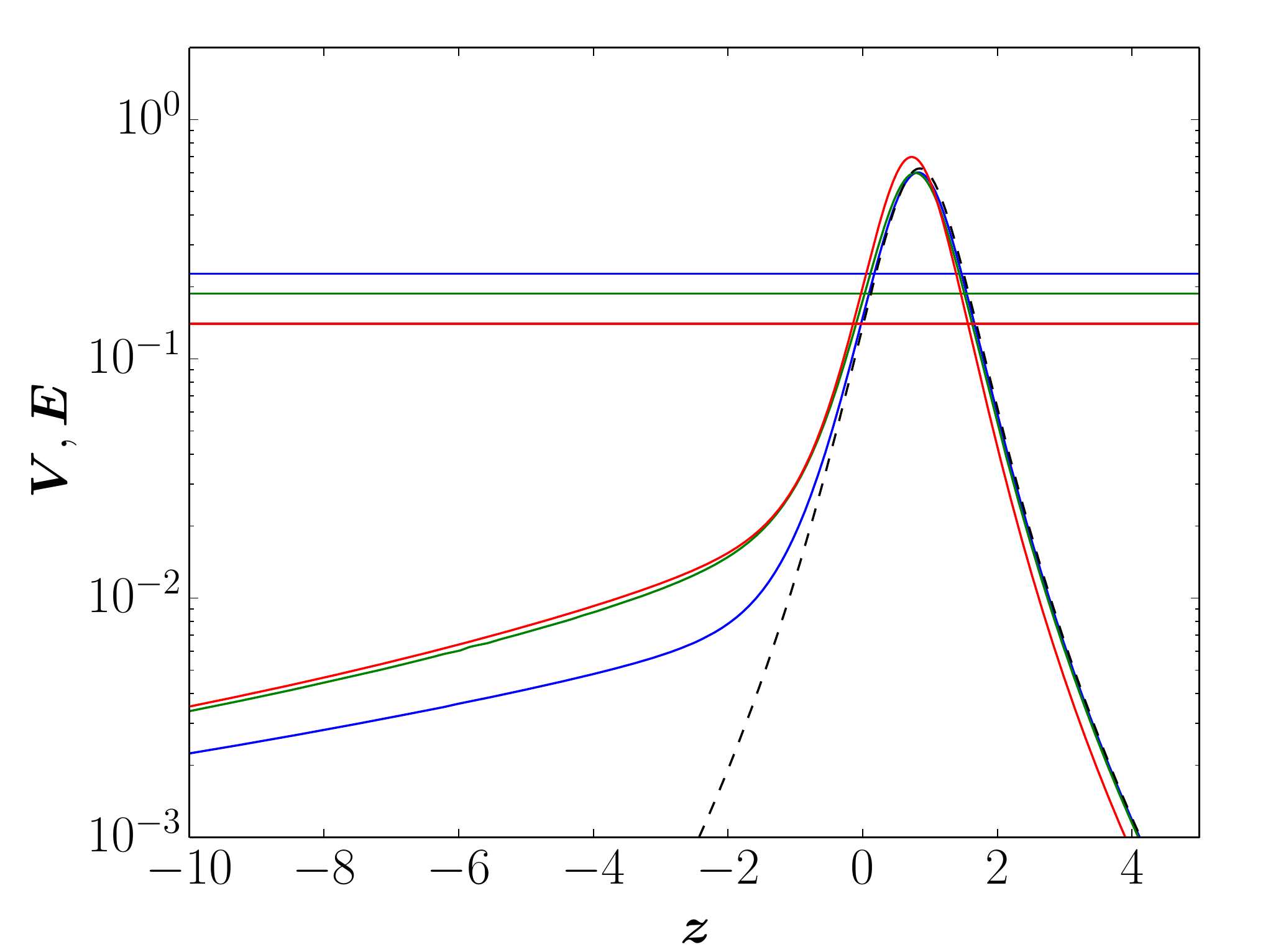}
  \end{center}
\caption{[Colors online] The plots represent the constants $\bE$
(horizontal lines) and the functions $\bV(\bz)$ (curves) calculated
for different scattering problems, corresponding to a fixed energy
$E=10^3 \Eunit$ and the CP potentials $V(z)$ for an hydrogen atom
above a perfect mirror, a silicon bulk and a silica bulk
(respectively blue, green and red from the highest to the lowest
value of $\bE$, or from the lowest to the highest value of $\bV$ in
the left-hand part of the plot). The dashed (black) curve is the
universal function $\bV_4(\bz)$. \label{fig:perfect-silicon-silica}
}
\end{figure}

In a similar manner, we can understand the dependence of QR
probabilities on the absolute magnitude of the CP potential. To do
so we consider hydrogen atoms falling onto a perfect mirror, a
silicon bulk or a silica bulk, which give rise respectively to
weaker and weaker CP interaction \cite{Dufour2013qrefl}.
Fig.\ref{fig:perfect-silicon-silica} shows the constants $\bE$ and
the functions $\bV(\bz)$ for a fixed energy $E=10^3 \Eunit \simeq
1.4$~neV. The potentials correspond to values for the far-end tails,
$C_4$ and $\ell$, which decrease from perfect mirror to silicon to
silica. As on Fig.\ref{fig:silica-energy}, the transformed
potentials $\bV$ have similar peak shapes, which tend to align on
the universal curve calculated for a pure $V_4$ potential and shown
as the dashed curve. In contrast, the transformed energies
$\bE=\kappa\ell$ decrease with $\ell$, which immediately explains
why the QR probability increases \cite{Dufour2013qrefl}.

We note that a similar discussion has been given in
\cite{Dufour2014liou} to explain the results obtained in
\cite{Dufour2013porous} for hydrogen atoms above nanoporous silica
with different porosities.

\section{Symmetry of the $V_4$ model}
\label{sec:C4model}

In this section, we discuss the model potential $V_4(z)=-C_4/z^4$
which is representative of the CP interaction in the far-end.
Furthermore, this model is interesting in itself because it obeys a
symmetry which enforces non trivial properties.

For the $V_4$ model, the WKB wave-vector has the simple form:
\begin{eqnarray}
\label{wkbkC4} &&k_\dB(z) = \sqrt{\kappa^2+\frac{\ell^2}{z^4}} ~.
\end{eqnarray}
This leads to a non trivial symmetry property for the Liouville
transformation corresponding to inversion:
\begin{equation}
\label{inversion} \tz = -\frac{\zeta^2}{z} ~,
\end{equation}
which maps the physical domain $z\in[0,\infty]$ into an inverted
domain $\tz \in[-\infty,0]$, while exchanging the roles of the
cliff-side and far-end.

The inversion \eqref{inversion} is an homographic function, so that
its Schwarzian derivative $\{z,\tz \}$ vanishes. If $\hz$ is another
map, chosen arbitrarily, Cayley's identity \eqref{cayley} leads to
$\left\{\hz,z\right\} = \left(\tz '(z)\right)^{2}\,\left\{\hz,\tz
\right\}$. When $\hz=\phi_\dB$, one deduces that the badlands
function defined as in \eqref{badlands} for the original and
inverted coordinates are identical:
\begin{equation}
 Q(z) = \frac{\{\phi_\dB,z\}}{2 k_\dB^2(z)} =
 \frac{\{\tphi_\dB ,\tz \}}{2 \tk_\dB ^2(\tz )}
 = \tilde Q (\tz ) ~,
\end{equation}
with $\tphi_\dB (\tz) \equiv \phi_\dB(z)$, $k_\dB \equiv
\phi_\dB'(z)$, $\tk_\dB \equiv \tphi_\dB'(\tz)$.

The badlands function may be written explicitly:
\begin{equation}
Q(z) = \frac{5 \kappa^2 \ell^2}{\left(\kappa^2 z^2 +
\frac{\ell^2}{z^2}\right)^3} = \frac{5 \kappa^2
\ell^2}{\left(\kappa^2 \tz ^2 + \frac{\ell^2}{\tz ^2}\right)^3} =
\tilde Q (\tz ) ~,
\end{equation}
and it reaches its peak value at $z=\zeta$, that is also $\tz
=\tilde \zeta = -\zeta$. This peak value scales as the inverse of
$\varkappa^2=\kappa\ell$. When multiplied by the latter value (see
\eqref{fixscale}), it leads to the universal function:
\begin{equation}
\label{universalV} \bV_4(\bz) = \frac{5}{\left(\frac{z^2}{\zeta^2} +
\frac{\zeta^2}{z^2}\right)^3} = \frac5{8\cosh^3(2u)} ~,\quad u
\equiv \ln\frac z\zeta ~,
\end{equation}
with a peak value $\tfrac58$ and a relation between $\bz$ and $u$
still to be discussed.

The WKB phase and the coordinate $\bz$ also obey symmetry properties
under the inversion:
\begin{eqnarray}
\label{universalz} \bz &=& \frac{\phi_\dB}\varkappa = \int_{u_0}^u
\sqrt{2\cosh(2u')} \, \D u' ~, \nonumber \\
&=& \bz_\ast + \int_{0}^u \sqrt{2\cosh(2u')} \, \D u' ~,
\end{eqnarray}
with $\bz_\ast$ the value corresponding to the inversion center:
\begin{eqnarray}
&& \bz_\ast \equiv \bz(\zeta) =
\frac{1}{\sqrt{\pi}}\Gamma\!\left(\tfrac34\right)^2  ~.
\end{eqnarray}
Eqs \eqref{universalV} and \eqref{universalz} constitute an explicit
parametric representation of the universal function $\bV_4(\bz)$
proving that it is symmetrical with respect to the inversion
$u\to-u$, that is also $\bz-\bz_\ast\to-\left(\bz-\bz_\ast\right)$.
Another representation in terms of hypergeometric functions is given
in the Appendix \ref{app:homogeneous}, where other homogeneous forms
of the potential $V(z)$ are also considered.

\begin{figure}[H]
  \begin{center}
   \includegraphics[width=3.2in]{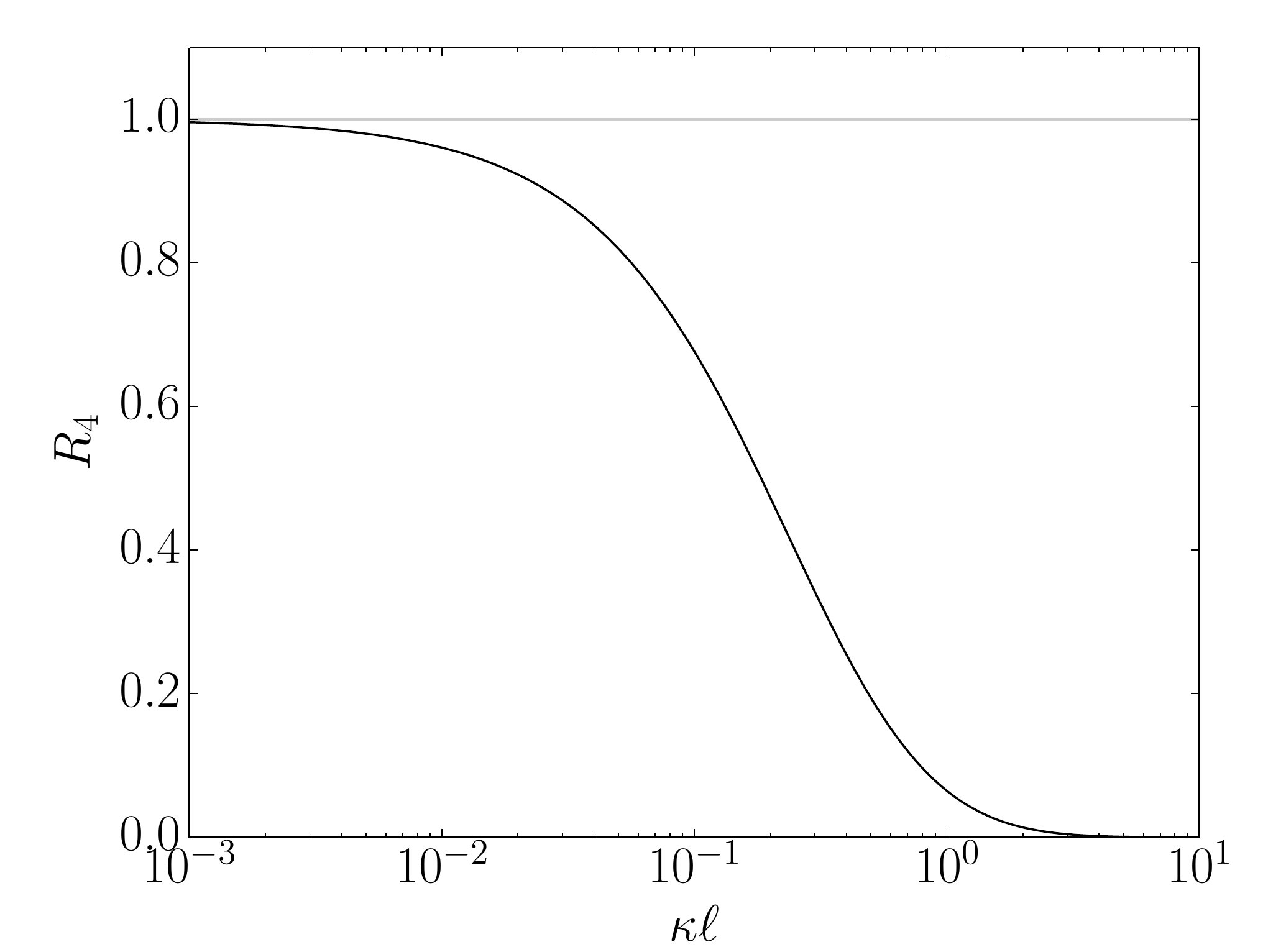}
  \end{center}
\caption{Quantum reflection probability $R_{4}$ calculated for the
$V_4$ model and shown as a function of the dimensionless parameter
$\kappa\ell$. \label{fig:QRC4}}
\end{figure}

The QR probability calculated for the $V_4$ model
\cite{OMalley1961,Gao2013}, denoted $R_{4}$ in the following and
plotted on Fig.~\ref{fig:QRC4}, is a universal function of the
dimensionless parameter $\kappa\ell$. It can be calculated by
numerically solving the Schr\"odinger equation for the potential
$V_4(z)$ or $\bV_4(\bz)$. Alternatively, it can be obtained by the
analytical method summed up in the Appendix \ref{app:analyticC4}.

\section{Discussion of QR probabilities}
\label{sec:discussion}

We now present the values obtained for QR probabilities, and compare
the exact results for the full CP potential with those obtained for
the $V_4$ model.

We first recall that the QR probability goes from unity at $\kappa
\to0$ to zero at $\kappa\to\infty$. Its departure from unity at low
energies is described by a scattering length $a$ defined by the
general relation:
\begin{equation}
\label{defa} r(\kappa) \underset{\kappa\to0}\simeq -\left(1 - 2
i\kappa a\right)~.
\end{equation}
The scattering length is a complex number, the imaginary part of
which determines the quantum reflection probability:
\begin{equation}
\label{defba} R(\kappa) \equiv \vert r(\kappa) \vert^2
\underset{\kappa \to0}\simeq 1-4 \kappa b \quad, \quad b\equiv
- \Im a~.
\end{equation}

Table \ref{bandell} gives $\ell$ and $b$ for an hydrogen atom above
a perfect mirror, a silicon bulk and a silica bulk, as obtained from
the full calculations in \cite{Dufour2013qrefl}. The table shows
that the equality $b=\ell$ typical of the $V_4$ model
\cite{Voronin2005} is no longer true for the full CP potential, with
large variations in particular for the case of silica bulks.

\begin{table}[H]
\begin{center}
\begin{tabular}{|c|c|c|c|}
\hline mirror & perfect & silicon & silica  \\
\hline $b$ [$a_0$] & 543.0 & 435.2 & 272.6 \\
\hline $\ell$ [$a_0$]  & 520.1 & 429.8 & 321.3\\
\hline
\end{tabular}
\caption{ Comparison of the values of $b$ and $\ell$ for an hydrogen
atom above a perfect mirror, a silicon bulk and a silica bulk, given
in atomic units $a_0\simeq52.9$~pm. \label{bandell}}
\end{center}
\end{table}

\begin{figure}[ht]
  \begin{center}
   \includegraphics[width=3.2in]{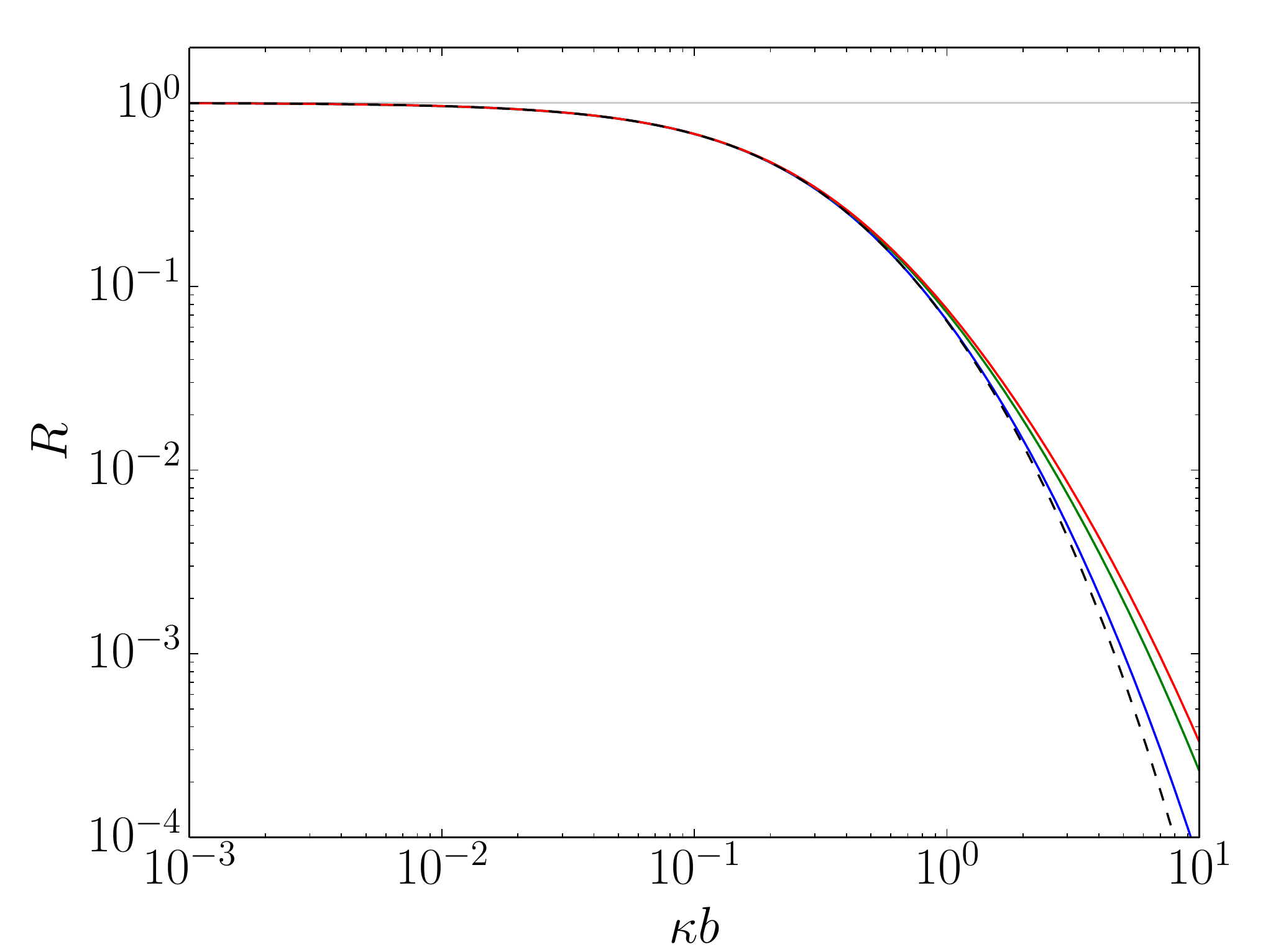}
  \end{center}
\caption{[Colors online] Log-log plot of the quantum reflection
probability $R$ shown as a function of the dimensionless parameter
$\kappa b$, corresponding respectively to a perfectly reflecting
mirrors (blue), silicon (green) and silica (red) bulks (from the
lowest to the highest curve at the right-hand side of the frame) and
compared to $R_4$ calculated for the $V_4$ model (black dotted
curve). \label{fig:Rvskbmats}}
\end{figure}

We show on Fig.\ref{fig:Rvskbmats} the calculated QR probabilities
$R$ as a function of the dimensionless parameter $\kappa b$ for the
scattering problems corresponding to Table \ref{bandell}. The full
curves represents the values calculated for perfect mirrors, silicon
and silica bulks \cite{Dufour2013qrefl}. They are compared to the
dashed curve which corresponds to the universal function $R_4$ (with
$b=\ell$ in this case) calculated for the pure $V_4$ model. Using
the value calculated for $b$ for different bulks of matter, it turns
out that the exact QR probabilities $R$ are close to the expression
$R_4$ evaluated for the same value of $\kappa b$. The agreement is
excellent at low energies because the peaks of $\bV$ on which QR
occurs correspond in these cases to values of $z$ in the far-end
tail of the CP potential. Differences between the curves $R$ and
$R_4$ appear at large enough energies, because distances $z$ closer
to the cliff-side play a significant role.

The same information is given in Table \ref{RandR4} with precise
numerical values of the QR probabilities $R$ calculated for the same
scattering problems at energy $E=10^3 \Eunit \simeq 1.407$~neV. This
corresponds to $\kappa=8.237\times10^{6}$~m$^{-1}$ that is also
$4.359\times10^{-4}\ a_0^{-1}$ (atomic units with
$a_0\simeq52.9$~pm). The comparison with $R_4(\kappa b)$ obtained
for the calculation of the $V_4$ model shows good agreement in
accordance with the fact that QR occurs in these cases in the
far-end tail of the CP potential. It is worth stressing that the
agreement would be much poorer when comparing $R$ to $R_4(\kappa
\ell)$.

\begin{table}[H]
\begin{center}
\begin{tabular}{|c|c|c|c|}
\hline mirror & conductor & silicon & silica  \\
\hline $\kappa b$ & 0.237 & 0.190 & 0.119 \\
\hline $R$ [\%] & 41.8 & 49.2 & 63.2 \\
\hline $R_4{(\kappa b)}$ [\%] & 41.9 & 49.0 & 63.1 \\
\hline
\end{tabular}
\caption{ Quantum reflection probabilities $R$ for hydrogen atoms
falling on a perfectly conducting, a silicon and a silica bulk
plate, with incident energy $E=10^3 \Eunit$, compared to the
corresponding value of the universal function $R_4(\kappa b)$. The
values of the $b$ and $\ell$ are given in Table \ref{bandell}.
\label{RandR4}}
\end{center}
\end{table}

In this paper, the problem of QR of an atom on a Casimir-Polder
attractive well has been mapped into an equivalent problem of
reflection on a wall through a Liouville transformation. This gauge
transformation  of the Schr\"odinger equation relates exactly
equivalent quantum scattering processes which correspond to
different semiclassical pictures. It produces a new interpretation
of the main features of quantum reflection and explains in a clear
manner the paradoxical features of the initial problem. It also
allows quantitative evaluation of QR probabilities which can be
obtained from the universal function corresponding to the pure $V_4$
model.

\medskip

\textit{Acknowledgements - } Thanks are due for insightful
discussions to M.-T. Jaekel, V.V. Nesvizhevsky, A. Yu. Voronin, and
the GBAR and GRANIT collaborations.

\appendix

\section{QR probability for the $V_4$ model}
\label{app:analyticC4}

In this appendix, we recall the analytical method which can be used
to to solve Schr\"odinger's equation for the $V_4$ model
\cite{OMalley1961,Gao2013}. The derivation presented here follows
the work of \cite{Holzwarth1973} and uses results in
\cite{Maclachlan1951,Olver2010}.

We first perform a Liouville transformation:
\begin{eqnarray}
z \to \tz(z)=\ln\frac z\zeta ~&,&\quad \npsi(z) \to \tpsi(\tz) =
\frac{\npsi(z)}{\sqrt{z}} ~.
\end{eqnarray}
With the new variables, the Schr\"odinger equation for the $V_4$
model becomes a modified Mathieu equation:
\begin{eqnarray}
\label{modmathieu} \tpsi''(\tz) + \left( -a + 2q \cosh(2\tz)
\right) \tpsi(\tz)=0 ~,
\end{eqnarray}
where $a\equiv\frac14$ while $q\equiv\varkappa=\sqrt{\kappa\ell}$ is
the only remaining parameter. A pair of solutions to this equation
can be written as series involving products of Bessel functions:
\begin{eqnarray}
\tpsi^{(\pm)}(\tz) = \sum_{n=-\infty}^\infty (-1)^n A_n^{(\tau)}
J_{\pm(n+\tau)}(\sqrt{q} e^{\tz}) J_{\pm n}(\sqrt{q}e^{-\tz}) ~.
\nonumber \\
\end{eqnarray}
Here $\tau$ is a complex parameter yet to be determined, known as
the Mathieu characteristic exponent. The coefficients $A_n^{(\tau)}$
obey the following recurrence relation:
\begin{eqnarray}
\left( (\tau+2n)^2 -a \right) A_n^{(\tau)} + q \left(
A_{n+1}^{(\tau)}+A_{n-1}^{(\tau)}  \right)=0 ~.
\end{eqnarray}
The infinite determinant associated with this system of equations
must be zero for a non trivial solution to exist. This singles out a
value of $\tau$, which can be obtained by following the procedure
detailed in \cite{Holzwarth1973} or using the Mathematica function
{\verb MathieuCharacteristicExponent[a,q] }. With the recurrence
relation we can write the ratios $A_n^{(\tau)}/A_{n-1}^{(\tau)}$ and
$A_{-n}^{(\tau)}/A_{-(n-1)}^{(\tau)}$ as continued fractions. These
ratios go to 0 when $|n|$ increases so that we can truncate the
continued fractions to obtain numerical values for $A_n^{(\tau)}$
(with $A_0^{(\tau)}=1$).

As a result of the invariance of equation \eqref{modmathieu} under
parity $\tz\to-\tz$ (which is the symmetry discussed in
\S~\ref{sec:C4model}), $\tpsi^{(\pm)}(-\tz)$ are also solutions, and
one can show that:
\begin{eqnarray}
\tpsi^{(\pm)}(\tz)=e^{\mp\sigma} \tpsi^{(\mp)}(-\tz)   ,\quad \sigma
= \ln \frac{\tpsi ^{(-)}(0)}{\tpsi ^{(+)}(0)} .
\end{eqnarray}
Using known results for the Bessel functions:
\begin{eqnarray*}
J_\nu(x)\underset{x\to \infty}{\simeq}\sqrt{\frac{2}{\pi
x}}\cos\left( x-\frac{\nu \pi}{2} - \frac{\pi}{4}\right)~,\quad
J_n(0)=\delta_{n,0}~,
\end{eqnarray*}
we deduce the asymptotic behaviors:
\begin{eqnarray}
\label{asymptot1} &&\tpsi ^{(\pm)}(\tz)
\underset{\tz\to\infty}{\simeq} \sqrt{\frac{2 }{\pi \sqrt{q}e^{\tz}
}} \cos\left(\sqrt{q}e^{\tz} \mp
\frac{\pi \tau}{2} -\frac{\pi}{4}  \right) ~,\\
&&\tpsi ^{(\pm)}(\tz) \underset{\tz\to-\infty}{\simeq} e^{\mp\sigma}
\sqrt{\frac{2 }{\pi \sqrt{q}e^{-\tz} }} \cos\left(\sqrt{q}e^{-\tz}
\pm \frac{\pi \tau}{2} -\frac{\pi}{4} \right) ~. \nonumber
\end{eqnarray}

As we are looking for the reflection and transmission amplitudes $r$
and $t$ for a wave coming from the far-end, we search the solution
$t\npsi_\lef^-(z)$ of Schr\"odinger equation \eqref{modmathieu}
which has the asymptotic behaviors:
\begin{eqnarray}
&& t\npsi_\lef^-(z) \underset{z\to0} {\simeq} \frac{tz}{\sqrt{\ell}}
\exp\left(-i \left(  2\varkappa\bz_\ast
- \frac{\ell}{z} \right) \right) ~, \nonumber \\
\label{asymptot2} && t\npsi_\lef^-(z) \underset{z\to\infty} {\simeq}
\frac{e^{-i \kappa z } + r e^{i \kappa z}}{\sqrt{\kappa}} ~.
\end{eqnarray}
We have used the asymptotic forms of $\phi_\dB(z)$:
\begin{eqnarray}
\phi_\dB(z)\underset{z\to0}{\simeq} 2 \varkappa \bz_*
-\frac{\ell}{z} ~, \quad \phi_\dB(z) \underset{z\to\infty}{\simeq}
\kappa z ~.
\end{eqnarray}

Matching the asymptotic forms (\ref{asymptot1}-\ref{asymptot2}), we
obtain the reflection and transmission amplitudes:
\begin{eqnarray}
&&r=-i \frac{\sinh(\sigma)}{\sinh(\sigma+i\pi\tau)}~,\quad
t=\frac{\sin(\pi \tau)
e^{2i\varkappa\bz_\ast}}{\sinh(\sigma+i\pi\tau)} ~.
\end{eqnarray}
It has been checked that this analytical method gives the same
results as a direct integration of the Schr\"odinger equation
\cite{Dufour2013qrefl}. The resulting QR probability $R_4=\vert
r\vert^2$ is drawn on Fig.~\ref{fig:QRC4}.

\section{Other homogeneous potentials}
\label{app:homogeneous}

In this appendix, we consider the case of homogeneous potentials
$V_n(z)=-C_n/z^n$ with $n>2$. This includes the already discussed
case $n=4$, as well as the case $n=3$ which corresponds to the Van
der Waals zone close to the surface and the case $n=5$ which
corresponds to the far-end of a slab mirror \cite{Dufour2013qrefl}.

We start by introducing two relevant length scales:
\begin{eqnarray}
&& \zeta_n=\sqrt[n]{\frac{C_n}E} ~,~  \ell_n = \sqrt[n-2]
{\frac{2mC_n}{\hbar^2}} = \sqrt[n-2] {\kappa^2(\zeta_n)^n} ~.
\end{eqnarray}
They generalize the definitions of $\zeta$ and $\ell$ for $n=4$.
$\zeta_n$ and $\ell_n$ measure respectively the distance at which
$E=\vert V_n\vert$ and the strength of the potential.

The WKB wavevector and phase are thus read:
\begin{eqnarray}
&&k_\dB(z)=\kappa \sqrt{1+\frac{1}{x^n} } ~,\quad x \equiv \frac z{\zeta_n} ~,\\
&&\phi_\dB(z)= \kappa\zeta_n \int_{x_0}^x \sqrt{1+\frac{1}{x'^n}}\
\D x' ~,
\end{eqnarray}
where $x_0$ is chosen to enforce \eqref{convz0}. For $n>2$, we
deduce:
\begin{eqnarray}
\phi_\dB &=& \frac{nx\kappa\zeta_n}{n-2} \\
&\times& \left(F\left(
\frac{1}{2},\frac{-1}{n};1-\frac{1}{n};\frac{-1}{x^n} \right) -
\frac{2}{n} \sqrt{1+\frac{1}{x^n}} \right) ~, \nonumber
\end{eqnarray}
where $F$ is the hypergeometric function defined as in
\cite{Olver2010}. The particular case $n=4$ gives an alternative
expression for the expressions in \S~\ref{sec:C4model}.

The badlands function:
\begin{eqnarray}
Q(z) = \frac{n x^{n-2}}{(\kappa \zeta_n)^2} \frac{4-n+4 (1+n) x^n}
{16 \left(1+x^n\right)^3}
\end{eqnarray}
is a peaked function reaching its maximum at:
\begin{eqnarray}
x_\ast = \sqrt[n] {\frac{5 n^2-3 n-8 + \sqrt{3(7 n^4-6 n^3-13
n^2)}}{4(n^2+3 n+2)}} ~.
\end{eqnarray}
We define the special Liouville gauge $\bz = \phi_\dB(z)/\varkappa$
as in \eqref{specific} and obtain $\bE=\varkappa^2$ and
$\bV=\varkappa^2 Q(z)$ as in \eqref{specificEV}. Up to know, we did
not fix the scale factor as in \eqref{fixscale}, as
$\ell\equiv\ell_4$ does plays not play any role for the potential
$V_n$ studied in this Appendix.

\begin{figure}[H]
  \begin{center}
    \includegraphics[width=3.2in]{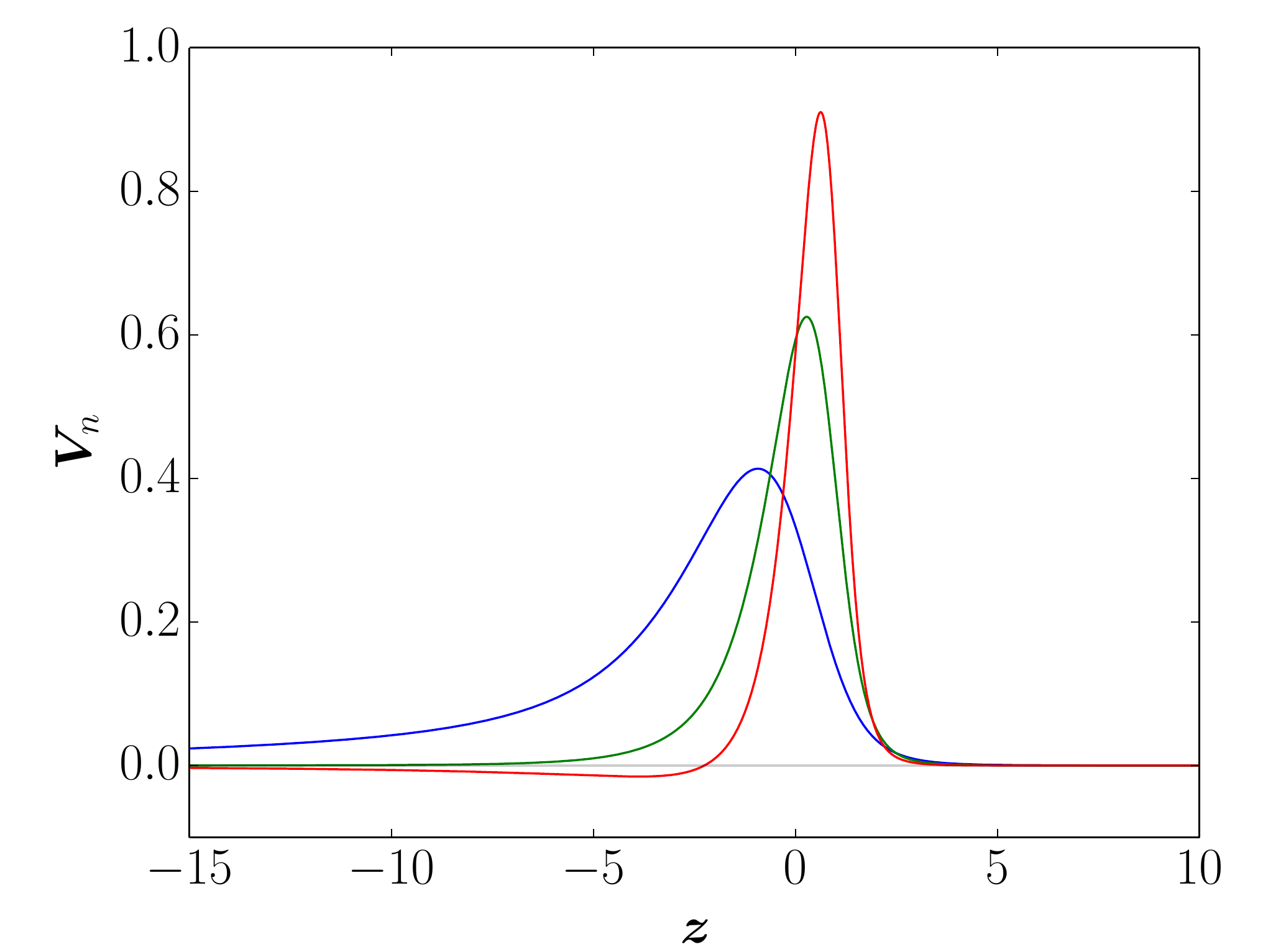}
  \end{center}
\caption{Plot of the universal functions $\bV_n(\bz)$ for $n=3,4,5$
(smallest to tallest). \label{fig:universalVn}}
\end{figure}

As the maximum value of $Q$ scales as $1/(\kappa \zeta_n)^2$, we
choose the scale factor as:
\begin{eqnarray}
\label{fixscalen} \varkappa_n = \kappa\zeta_n ~,
\end{eqnarray}
which generalizes the definition \eqref{fixscale}. This leads to
universal functions $\bV_n$ which do not depend on any other
parameter than $n$. The functions $\bV_n(\bz)$ are drawn on
Fig.\ref{fig:universalVn} for the cases $n=3,4,5$, as functions of
the coordinate $\bz = \phi_\dB(z)/\varkappa$ using the convention
\eqref{fixscalen}.

With the same convention, the integrals \eqref{integral} of $\bV_n$
over the real axis are real numbers depending only on $n$ and they
can be expressed in terms of the Gamma function:
\begin{eqnarray}
\bI_n = \frac{n\sqrt{\pi } \Gamma\left(2+\frac{1}{n}\right)
\sec\left(\frac{\pi }{n}\right)}{12
\Gamma\left(\frac{1}{2}+\frac{1}{n}\right)} ~.
\end{eqnarray}
In particular, the case $n=4$ corresponds to:
\begin{eqnarray}
\bI_4=\frac{5 \Gamma(5/4)^2}{3 \sqrt{\pi } } \simeq 0.772531 ~.
\end{eqnarray}

\bibliography{liouvilletrans.bbl}

\end{document}